\documentclass[10pt, journal]{IEEEtran}

\usepackage[T1]{fontenc}
\usepackage[english]{babel}	
\usepackage[latin1]{inputenc}

\usepackage{amsmath,
	amssymb,
	amsthm}

\usepackage{balance}
\usepackage[noadjust]{cite}

\usepackage{color}
\usepackage{graphicx}
\graphicspath{{Figures/}}

\usepackage{pgfplots}
\pgfplotsset{compat=1.12,
	legend style={font=\footnotesize},
}

\makeatletter
\let\MYcaption\@makecaption
\makeatother

\usepackage[font=footnotesize]{subcaption}

\makeatletter
\let\@makecaption\MYcaption
\makeatother

\usepackage{ctable}

\colorlet{green}{green!60!black}
\IEEEoverridecommandlockouts

\newtheorem{example}{Example}

\usepackage{tikz}
\usetikzlibrary{calc}

\newcommand{\cost}{\beta}
\newcommand{\updcost}{\cost_\text{C}}

\newcommand{\sbscost}{\cost_\text{SBS}}
\newcommand{\updaterate}{L_\text{C}}
\newcommand{\mbsrate}{L_\text{MBS}}
\newcommand{\sbsrate}{L_\text{SBS}}

\newcommand{\Rsbs}{R_\text{SBS}}
\newcommand{\Rupd}{R_\text{upd}}
\newcommand{\Rmem}{R_\text{mem}}

\renewcommand{\S}{\mathcal{S}}
\newcommand{\A}{\mathcal{A}}

\newcommand{\boldtheta}{\boldsymbol{\theta}}
\newcommand{\boldphi}{\boldsymbol{\phi}}

\newcommand{\boldmu}{\boldsymbol{\mu}}
\newcommand{\boldmubar}{\bar\boldmu}

\newcommand{\refagent}{b_\text{ref}}

\newcommand{\B}{\mathcal{B}}
\newcommand{\buffer}{\mathcal{M}}
\newcommand{\replay}{\mathcal{T}}

\newcommand{\xbar}{\bar{x}}
\newcommand{\mubar}{\bar{\mu}}

\renewcommand{\d}{\text{d}}
\newcommand{\E}{\mathbb{E}}
\newcommand\floor[1]{\left\lfloor#1\right\rfloor}
\DeclareMathOperator*{\argmax}{arg\,max}

\title{Dynamic Coded Caching in Wireless Networks\\Using Multi-Agent Reinforcement Learning}

\begin{document}

\author{Jesper Pedersen, Alexandre Graell i Amat,~\IEEEmembership{Senior Member,~IEEE},\\ Fredrik Br\"annstr\"om,~\IEEEmembership{Member,~IEEE}, and Eirik Rosnes,~\IEEEmembership{Senior Member,~IEEE}
\thanks{This work was funded by the Swedish Research Council under grant 2016-04253.}
\thanks{J. Pedersen, A. Graell i Amat, and F. Br\"annstr\"om are with the Department of Electrical Engineering, Chalmers University of Technology, SE-41296 Gothenburg, Sweden (e-mail: \{jesper.pedersen, alexandre.graell, fredrik.brannstrom\}@chalmers.se).}
\thanks{E. Rosnes is with Simula UiB, N-5020 Bergen, Norway (e-mail: eirikrosnes@simula.no).}
\vspace{-3ex}}

\maketitle

\begin{abstract}
We consider distributed caching of content across several small base stations (SBSs) in a wireless network, where the content is encoded using a maximum distance separable code.
Specifically, we apply soft time-to-live (STTL) cache management policies, where coded packets may be evicted from the caches at periodic times. We propose a reinforcement learning (RL) approach to find coded STTL policies minimizing the overall network load. We demonstrate that such caching policies achieve almost the same network load as policies obtained through optimization, where the latter assumes perfect knowledge of the distribution of times between file requests as well the distribution of the number of SBSs within communication range of a user placing a request. We also suggest a multi-agent RL (MARL) framework for the scenario of non-uniformly distributed requests in space. For such a scenario, we show that MARL caching policies achieve lower network load as compared to optimized caching policies assuming a uniform request placement. We also provide convincing evidence that synchronous updates offer a lower network load than asynchronous updates for spatially homogeneous renewal request processes due to the memory of the renewal processes.
\end{abstract}

\begin{IEEEkeywords}
	Caching, content delivery networks, deep deterministic policy gradient, erasure correcting codes, multi-agent reinforcement learning, time-to-live.
\end{IEEEkeywords}

\section{Introduction}
Caching is a promising technology to reduce the amount of data transmitted over congested or latency-prone backhaul links, effectively trading affordable memory for expensive bandwidth resources \cite{Boccardi2014}. Frequently requested content may be cached in small base stations (SBSs) or smart devices with spare memory capacity. Users can then download requested content directly from SBSs, or from caching devices using device-to-device (D2D) communication. For the case when content is cached in a distributed fashion, the use of erasure correcting codes (ECCs) has been shown to decrease the amount of data that has to be fetched over the backhaul \cite{Shanmugam2013, Bioglio2015, Pedersen2016, Pedersen2019}. Furthermore, caching facilitates index-coded multicasts, which may greatly reduce the amount of data that has to be transmitted to the end-users \cite{Maddah-Ali2014}. Index coding has also been used in conjunction with ECCs to improve performance in D2D caching networks \cite{Ji2016}. All the aforementioned papers consider the cached content to remain static for an extended period of time (e.g., one day) according to a given file popularity distribution.

An alternative to static caching is to responsively cache a requested file and maintain it in the cache for a limited time, an approach referred to as time-to-live (TTL) caching. TTL caching reduces backhaul traffic under a renewal request process, i.e., a process with memory \cite{Pedersen2020}. The soft-TTL (STTL) caching policies suggested by Goseling and Simeone \cite{Goseling2019}, where fractions of files may be evicted from the cache at periodic times, were shown to further reduce the backhaul traffic. In \cite{Pedersen2020}, we extended the STTL scheme to consider the periodic eviction of coded packets in a distributed caching scenario. We demonstrated that distributed coded STTL can significantly reduce the backhaul traffic, in particular when the request process is bursty \cite{Pedersen2020}. The analysis in \cite{Pedersen2020} relies on a number of important assumptions to make the problem tractable. First, in order to provide optimal coded caching policies, adhering to a long-term average cache size constraint, both the request process statistics as well as the steady-state distribution of the number of SBSs accessible to a user placing a request have to be known. Secondly, the caches are assumed to be synchronously updated, i.e., all caches store always the same amount of each file. The assumptions of perfectly known request statistics and SBS distribution might not be realizable in practice, or the estimates thereof might be inaccurate. Furthermore, synchronous cache updates is not necessarily optimal under a spatially non-uniform request process, and it remains unclear whether always updating all caches at the same time is optimal at all.

In this paper, we reformulate the distributed coded STTL caching problem in \cite{Pedersen2020}, where the caches are synchronously updated, as a reinforcement learning (RL) problem, with the benefit of not requiring any prior knowledge of request statistics or SBS distribution. In particular, we utilize the deep deterministic policy gradient (DDPG) algorithm, where the action-value and policy functions are approximated by neural networks \cite{Lillicrap2019}. We show that the proposed RL algorithm can obtain STTL caching policies achieving almost the same reduction in backhaul traffic as the optimal policies in \cite{Pedersen2020}, where the latter assumes perfect knowledge of request statistics and SBS distribution. We then lift the assumption of synchronized cache updates and devise a multi-agent RL (MARL) algorithm in order to study the backhaul traffic reductions when SBSs are autonomously deciding on caching policies. Our results suggest that synchronous cache updates lead to larger reductions in backhaul traffic as compared to asynchronous updates under a spatially uniform renewal request process, i.e., the fact that the request process has memory has a bigger impact on the backhaul traffic than the cost of updating all SBSs synchronously. Finally, we show that the caching policies obtained using the MARL approach outperform the policies acquired using the optimization framework in \cite{Pedersen2020}, where the latter assumes synchronously updated caches, for spatially non-uniform request processes.

\subsection{Related Work}
Several papers pursue an RL approach to find the file popularity profile and decide what content to cache in order to minimize some cost function, e.g., sum delay, rate, energy, or data, due to content downloads. In \cite{Blasco2014}, the file popularity is found using a multi-armed bandit (MAB) formulation which aids the decision on the content allocation in a single cache. A similar approach is considered in \cite{Song2017} for content caching in multiple SBSs. The caching of files from a dynamic library was considered in \cite{Somuyiwa2018}, a time-varying popularity profile was studied in \cite{Sadeghi2018}, and the distributed caching of content in mobile devices was investigated in \cite{Jiang2019}. All the aforementioned works consider the caching of uncoded content.

The deployment of RL techniques to optimize the caching of coded content was initially proposed in \cite{Sengupta2014}. Specifically, the file popularity profile is learned using a MAB framework and the coded content placement is optimized separately. Instead, \cite{Gao2020} uses an RL approach, specifically Q-learning, for a unified approach that both estimates the file popularity and makes decisions on coded content allocations.
The caching of coded content, assuming an unknown and time-varying file popularity profile, is considered in \cite{Zhang2021}. Furthermore, the pre-training of actor and critic networks is suggested to expedite the learning. Finally, the case where SBSs autonomously decide on coded caching policies is analyzed in \cite{Wu2020}, using an MARL approach.

The papers \cite{Blasco2014,Song2017,Somuyiwa2018,Sadeghi2018,Jiang2019,Sengupta2014,Gao2020,Zhang2021,Wu2020} estimate request statistics and make decisions on static content allocations. After a period of time the process is repeated with a new estimation and content allocation decision. Our work is fundamentally different in that the caching of coded content is instead triggered by a file request. Coded packets may subsequently be evicted from the cache at periodic times after the last request, something we observed can drastically increase the amount of data that can be downloaded from the caches under a renewal request process \cite{Pedersen2020}.

\subsection{Notation}
For a random variable $X$, $\sim$ indicates the distribution of $X$ and the expectation of $X$ is denoted by $\E[X]$. We use calligraphic letters, e.g., $\B$, to denote a set. The cardinality of a set $\B$ is denoted by $|\B|$. We use the notation $\B \backslash b$ for the set $\B$ excluding the element $b \in \B$. Finally, vectors are denoted in bold, e.g., $\boldtheta$.

\section{System Model}
\label{sec:model}
\begin{figure}
	\centering
	\includegraphics[width=\columnwidth]{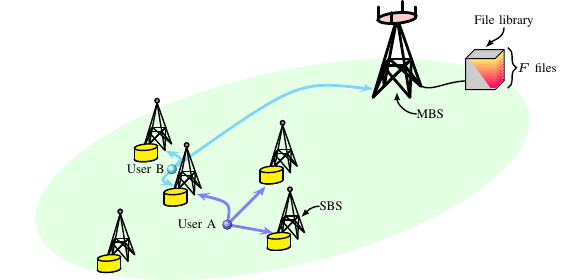}
	\caption{The system model with an MBS, with access to the file library consisting of $F$ files, a number of SBSs, and users A and B. User A downloads coded packets from a number of SBSs and can decode the requested file. User B also downloads coded packets from SBSs but have to download additional coded packets from the MBS to decode the file.}
	\label{fig:model}
\end{figure}

We consider an area served by a macro base station (MBS) with access to a file library of $F$ files. All files $f = 1, 2, \ldots, F$ are of equal size and, without loss of generality, we assume that the file size is normalized to $1$. $B$ SBSs are deployed in the area, each with storage capacity equivalent to $C$ files. Mobile users request files from the library at random times, where we denote the rate at which file $f$ is requested by $\omega_f$. Furthermore, we let $p_f = \omega_f / \omega$, where $\omega$ is the sum request rate in the area, i.e., $\sum_{f=1}^F \omega_f = \omega$. We assume that users can download requested files from an SBS if it is within range $r$ and we denote by $\B^{(t)}$ the set of SBSs within communication range of a user placing a request at a given time $t$. The system model considered in this paper is shown in Fig.~\ref{fig:model}.

\subsection{Caching Policy}
\label{sec:policy}
Each file to be cached is partitioned into smaller packets and encoded using a maximum distance separable (MDS) code. In response to a request for file $f$, coded packets of this particular file are cached and may subsequently be evicted from the cache over time. The cached coded packets correspond to a fraction cached of file $f$. Specifically, we denote by $x_{f, b}(\tau)$, $0 \le x_{f, b}(\tau) \le 1$, the fraction of file $f$ cached by SBS $b$ at time $\tau$ since the most recent request for the file, i.e., $\tau$ is the \emph{inter-request time}.
Similar to \cite{Goseling2019, Pedersen2020}, we refer to $x_{f, b}(\tau)$ as the \emph{caching policy}. The STTL caching policy in \cite{Goseling2019}, extended in \cite{Pedersen2020} to a distributed coded caching scenario, dictates that coded packets can be evicted at periodic times with period $T$ after the most recent request for file $f$ and that a total of $K$ cache updates are allowed. Hence, the caching policies are
\begin{equation}
\label{eq:policy}
	x_{f, b}(\tau) = \begin{cases}
		x_{f, b}^{(j)}, & \text{if}~jT \le \tau < (j+1)T, j = 0, \ldots, K-1,\\
		x_{f, b}^{(K)}, & \text{if}~\tau \ge KT,
	\end{cases}
\end{equation}
where
\begin{equation}
\label{eq:decpol}
	1 \ge x_{f, b}^{(0)} \ge x_{f, b}^{(1)} \ge \ldots \ge x_{f, b}^{(K)} \ge 0.
\end{equation}
The left inequality follows from the fact that we never cache more than the size of the file. Note that, by the MDS property \cite{MacWilliams1977}, obtaining this fraction of data is sufficient to decode the file. The subsequent inequalities prescribe that coded packets may be evicted over time. We will refer to $jT \le \tau < (j+1) T$ as \emph{time-slot} $j = 0, 1, \ldots, K-1$ and to $\tau \ge KT$ as time-slot $K$. We stress that the caching policies $x_{f, b}(\tau)$ dictate how the caches are replenished at the time of a request for a particular file and how cached fractions (i.e., coded packets) are subsequently evicted over time, i.e., as a function of the inter-request time.

Note that \eqref{eq:decpol} implies that coded packets can only be evicted from the cache. Here, we lift this assumption to make the problem more tractable to our RL framework. Specifically, by lifting \eqref{eq:decpol}, we assume that coded packets can be \emph{added} or evicted at periodic times, i.e.,
\begin{equation}
\label{eq:incdecpol}
	0 \le x_{f, b}^{(j)} \le 1,~j = 0, 1, \ldots K,~b = 1, 2, \ldots B.
\end{equation}
Hence, we study caching policies that are more general than STTL. However, the connection between the caching policy \eqref{eq:policy} and the hazard function $h(\tau)$, measuring the probability to observe a request a time $\tau$ after the previous request, was demonstrated in \cite{Goseling2019}. Specifically, for non-increasing $h(\tau)$, the caching policy \eqref{eq:policy} will be non-increasing in $j$. Consequently, a caching policy adhering to the constraint \eqref{eq:incdecpol} will also abide by the constraint \eqref{eq:decpol}.

\subsection{Content Download}
A user placing a request at time $t$ for file $f$ downloads coded packets from the SBSs within communication range. Let $\mu_{f, b}^{(t)}$, $0 \le \mu_{f, b}^{(t)} \le 1$, denote the amount of file $f$ that is cached at time $t$ at SBS $b \in \B^{(t)}$. Note that, in contrast to the caching policy $x_{f, b}(\tau)$ describing the eviction of coded packets over time in relation to the inter-request time, $\mu_{f, b}^{(t)}$ is the realized cached amount due to the caching policy and the inter-request time. The user placing the request downloads as much data as possible from the SBSs (up to $1$, the size of the file). For an MDS code, the file can be decoded if coded packets corresponding to an amount
\begin{equation}
\label{eq:sufdec}
	\sum_{b \in \B^{(t)}} \mu_{f, b}^{(t)} \ge 1
\end{equation}
are retrieved \cite{MacWilliams1977}. If additional coded packets are required to decode the file, i.e., if
\begin{equation}
	\sum_{b \in \B^{(t)}} \mu_{f, b}^{(t)} < 1,
\end{equation}
these packets are downloaded from the MBS. Consequently, a user placing a request for file $f$ downloads an amount
\begin{equation}
\label{eq:sbsdata}
	\min \left\{ \sum_{b \in \B^{(t)}} \mu_{f, b}^{(t)}, 1 \right\}
\end{equation}
from SBSs within communication range, where the $\min\{ \cdot, \cdot \}$ function prevents more data to be downloaded than is sufficient to decode the requested file (see \eqref{eq:sufdec}). Furthermore, the user downloads coded packets from the MBS corresponding to an amount
\begin{equation}
\label{eq:mbsdata}
	\max \left\{ 1 - \sum_{b \in \B^{(t)}} \mu_{f, b}^{(t)}, 0 \right\}.
\end{equation}
The following example illustrates the partitioning of a file $f$, the encoding using an MDS code, and the caching policy dictating the number of coded packets cached and evicted over time presented in Section~\ref{sec:policy} as well as the amount downloaded from SBSs and the MBS.

\begin{example}
	Consider $B = 2$ SBSs deployed in the area, an update period $T = 1$, $K = 2$ cache updates, and the caching policies
	\begin{equation}
		x_{f, 1}^{(j)} = \begin{cases}
				1/2, & \text{if}~j = 0,\\
				0, & \text{if}~j = 1, 2
			\end{cases}
	\end{equation}
	and
	\begin{equation}
		x_{f, 2}^{(j)} = \begin{cases}
				1, & \text{if}~j = 0,\\
				2/3, & \text{if}~j = 1,\\
				1/3, & \text{if}~j = 2.
			\end{cases}
	\end{equation}
	To realize these caching policies, we partition the file $f$ into $6$ packets, each of size $1/6$, and encode them into $9$ coded packets, also of size $1/6$, using an MDS code.
	
	For simplicity, we consider a request for file $f$ at time $t = 0$ and assume that the user placing the request is within communication range of both SBSs. Hence, at time $t = 0$, SBS $b = 1$ caches $3$ coded packets corresponding to a fraction $3/6 = 1/2$ of file $f$. Similarly, SBS $b = 2$ caches $6$ coded packets corresponding to a fraction $6/6 = 1$ of file $f$. We assume that there is no request after a time $t = T = 1$. Consequently, $3$ and $2$ coded packets are evicted from SBS $b = 1$ and $b = 2$, respectively. If there is no request for the file until a time $t = 2T = 2$, SBS $b = 2$ evicts $2$ coded packets from the cache and consequently caches a fraction $2/6 = 1/3$.
	
	Assume that, at time $t = 2.6$, there is another request for file $f$ and that $\B^{(2.6)} = \{1, 2\}$. At $t = 2.6$, SBS $b = 1$ no longer caches any coded packets of file $f$ and SBS $b = 2$ still caches a fraction $1/3$ of the file. Consequently, using \eqref{eq:sbsdata}, the user requesting the file downloads an amount $1/3$ from SBS $b = 2$ and, using \eqref{eq:mbsdata}, downloads an amount $2/3$ from the MBS.
\end{example}

To retrieve the file a user will (in general) download data from SBSs and the MBS. Each file $f$ is requested with rate $\omega_f$ (in requests per time), which corresponds to an amount of data (i.e., coded packets) of file $f$ downloaded from the MBS and SBSs per time. 
We refer to the sum (over files) of data downloaded per time from the MBS as the MBS download rate, and denote it by $\mbsrate$. Similarly, we define the SBS download rate and denote it by $\sbsrate$. We also consider the amount of data sent to update the caches. Similar to the download rate, we refer to the amount of data sent to update the caches per time as the update rate and denote it by $\updaterate$. (See \cite[Section~IV]{Pedersen2020} for an explicit derivation of $\mbsrate$, $\sbsrate$, and $\updaterate$ for the case of synchronously updated caches as well as known request statistics and SBS distribution.) Furthermore, we define the weighted sum of link rates as the \emph{network load},
\begin{equation}
\label{eq:netwload}
	L \triangleq \mbsrate + \sbscost \sbsrate + \updcost \updaterate
\end{equation}
where $\sbscost \le 1$ and $\updcost \le 1$ are the costs, representing, e.g., delay or energy, relative to the cost of downloading data from the MBS.

\subsection{Problem Formulation}
The goal is to optimize the caching policies \eqref{eq:policy} to minimize the network load \eqref{eq:netwload}, without requiring prior knowledge of request statistics and SBS distribution. Using \eqref{eq:sbsdata} and \eqref{eq:mbsdata}, we note that minimizing the network load corresponds to maximizing
\begin{equation}
\label{eq:premax}
	(1 - \sbscost) \sbsrate - \updcost \updaterate,
\end{equation}
which was also shown in \cite[Section~IV]{Pedersen2020}. Maximizing \eqref{eq:premax} depends on the relative cost
\begin{equation}
\label{eq:relcost}
	\frac{\updcost}{1 - \sbscost}\;.
\end{equation}
Therefore, without loss of generality, we can set $1-\sbscost= 1$ and vary $\updcost$.
Hence, obtaining caching policies minimizing \eqref{eq:netwload} is equivalent to find policies maximizing
\begin{equation}
\label{eq:maximization}
	\sbsrate - \updcost \updaterate.
\end{equation}
To obtain caching policies maximizing \eqref{eq:maximization}, without requiring prior knowledge of request statistics and SBS distribution, we use RL techniques, which are introduced in the next section.

\section{Preliminaries}
\label{sec:prel}
In this section, we briefly introduce Markov decision processes (MDPs) as well as the DDPG algorithm used to solve them. An MDP is a mathematical framework used to model sequential decision making. In RL, we describe the entity taking actions as the agent and the agent interacts with an environment through observing states and rewards due to particular actions \cite[Chapter~3]{Sutton2018}. At discrete time-steps, $t = 0, 1, 2, \ldots$, the agent observes \emph{state} $S_t \in \S$ and consequently takes \emph{action} $A_t \in \A$, where $\S$ and $\A$ are the sets of possible states and actions, also referred to as the state and action spaces, respectively. In response to the action, the agent observes a new state $S_{t+1} \in \S$ and receives a \emph{reward} $R_{t+1}$ and the process continues. The \emph{discounted return} is defined as
\begin{equation}
	G_t \triangleq \sum_{k=0}^\infty \gamma^k R_{t+k+1},
\end{equation}
where $\gamma$, $0 \le \gamma \le 1$, is the discount rate \cite[Section~3.3]{Sutton2018}. A policy is a mapping from states to actions \cite[Section~3.5]{Sutton2018}. For a \emph{stochastic policy}, we specify the probability
\begin{equation}
	\pi(a|s) \triangleq \Pr \{ A_t = a | S_t = s \}
\end{equation}
of taking action $a$ given that the state is $s$. The \emph{action-value function} $q_\pi(s, a)$ expresses the expected discounted return of being in a state $s$, taking action $a$, and following the policy $\pi$ afterward, i.e., \cite[Section~3.5]{Sutton2018}
\begin{equation}
	q_\pi(s, a) \triangleq \E_\pi \left[ G_t | S_t = s, A_t = a \right].
\end{equation}
Our goal is to find the optimal action-value function $q_*(s, a)$\textemdash equivalently, the optimal policy\textemdash such that \cite[Section~3.6]{Sutton2018}
\begin{equation}
	q_*(s, a) = \max_\pi q_\pi(s,a).
\end{equation}
The model-free Q-learning algorithm \cite{Watkins1989} is widely used to acquire optimal policies. However, this algorithm requires that the state and action spaces are finite. For continuous state and action spaces, as it is the case for the considered caching problem, we must resort to other methods. The DDPG algorithm suggested in \cite{Lillicrap2019} is well suited for the problem at hand. In the following, we briefly introduce this algorithm.

\subsection{Deep Deterministic Policy Gradient}
A \emph{deterministic policy} $\eta(s)$ maps states to actions, i.e., $A_t = \eta(s)$, if $S_t = s$ \cite{Lillicrap2019}. We parametrize the policy by means of a neural network (NN) with weights $\boldtheta$, referred to as the \emph{actor} network, and denote it by $\eta_{\boldtheta} (s)$. Similarly, we parametrize the action-value function by a NN with weights $\boldphi$, referred to as the \emph{critic} network, denoted by $q_{\boldphi} (s, a)$. The aim of the DDPG algorithm is to learn the weights $\boldtheta$ such that $\eta_{\boldtheta} (s)$ approximates well the best action chosen from the optimal action-value function, i.e., \cite{Lillicrap2019}
\begin{equation}
	\eta_{\boldtheta} (s) \approx \argmax_a q_* (s, a).
\end{equation}
To maintain exploration, Gaussian noise is added to the (continous) action, i.e.,
\begin{equation}
	\eta_{\boldtheta} (s) + \epsilon
\end{equation}
where $\epsilon \sim \mathcal{N}(0, \sigma^2)$. Simultaneously, the algorithm attempts to learn the weights $\boldphi$ such that $q_{\boldphi} (s, a)$ approximates well the optimal action-value function, i.e.,
\begin{equation}
	q_{\boldphi} (s, a) \approx q_* (s, a).
\end{equation}
In order to stabilize the training, two corresponding \emph{target} networks are maintained, a target actor network with weights $\tilde\boldtheta$ parametrizing the deterministic policy $\eta_{\tilde\boldtheta}(s)$ and a target critic network with weights $\tilde\boldphi$ parametrizing the action-value function $q_{\tilde\boldphi}(s, a)$ \cite{Lillicrap2019}.
To further stabilize the training, tuples $(S_t = s, A_t = a, S_{t+1} = s', R_{t+1} = r, d)$ of states, actions, state transitions, and rewards previously experienced by the algorithm, where $d = 1$ if $s'$ is a terminal state and $d = 0$ otherwise, are stored in a \emph{memory replay buffer}, denoted by $\buffer$ \cite{Lillicrap2019}. For a set of tuples $\replay = \{(s, a, s', r, d)\}$ sampled randomly from $\buffer$, the weights of the critic network are updated by stochastic gradient descent,
\begin{equation}
	\nabla_{\boldphi} \frac{1}{|\replay|} \sum_{(s, a, s', r, d) \in \replay} ( q_{\boldphi}(s, a) - y(r, s', d) )^2,
\end{equation}
where
\begin{equation}
	y(r, s', d) \triangleq r + \gamma (1-d) q_{\tilde\boldphi} \left( s', \eta_{\tilde\boldtheta}(s') \right).
\end{equation}
The weights of the actor network are modified by gradient ascent,
\begin{equation}
	\nabla_{\boldtheta} \frac{1}{|\replay|} \sum_{s \in \replay} q_{\boldphi}(s, \eta_{\boldtheta}(s)),
\end{equation}
where, with slight abuse of notation, $s \in \replay$ refers to $s$ extracted from the tuple $(s, a, s', r, d) \in \replay$. The target network weights are updated as
\begin{align}
	\tilde\boldtheta & \longleftarrow \rho \tilde\boldtheta + (1 - \rho) \boldtheta,\\
	\tilde\boldphi & \longleftarrow \rho \tilde\boldphi + (1 - \rho) \boldphi,
\end{align}
where $\rho$ is the target network update parameter, which is close to $1$. The algorithm is hence attempting to simultaneously minimize an expression closely resembling the Bellman optimality equation for the action-value function \cite[Section~3.6]{Sutton2018} by modifying the critic network weights $\boldphi$, while maximizing the action-value function by adjusting the actor weights $\boldtheta$ \cite{Lillicrap2019}.

\section{RL-Based Coded Caching for\\Synchronous Cache Updates}
\label{sec:synch}
In this section, we consider the case where caches are synchronously updated, which is a reasonable choice under spatially homogenous request processes as file requests are equally likely to be within communication range of any SBS. For synchronous updates, all caches are updated despite of which SBSs are within communication range of the user placing the request for a file. Furthermore, the caching policy is the same for all SBSs, i.e.,
\begin{equation}
\label{eq:synchpol}
	x_{f, 1}^{(j)} = x_{f, 2}^{(j)} = \ldots = x_{f, B}^{(j)} = x_{f}^{(j)},~\text{for all}~f, j,
\end{equation}
in \eqref{eq:policy}. As a consequence, all SBSs cache always the same amount of each file, although the cached data correspond to unique coded packets,
\begin{equation}
\label{eq:synch}
	\mu_{f, 1}^{(t)} = \mu_{f, 2}^{(t)} = \ldots = \mu_{f, B}^{(t)} = \mu_f^{(t)},~\text{for all}~f, t.
\end{equation}

For this scenario, we propose an RL approach to determine good coded STTL caching policies \eqref{eq:policy} (using \eqref{eq:synchpol}) to maximize \eqref{eq:maximization}, equivalently minimizing the network load \eqref{eq:netwload}. In contrast to the optimization of the caching policies for synchronous cache updates in \cite{Pedersen2020}, which requires perfect knowledge of the request statistics and SBS distribution, the proposed RL approach requires neither. We define the problem as an MDP and apply the DDPG algorithm (see Section~\ref{sec:prel}) to learn optimal caching policies. Before defining the MDP formally in Sections~\ref{sec:synch}-A--E, we offer a general description.

We assume that the RL agent observes file requests sequentially, where we model each request as a time-step in the MDP. In addition to the presently requested file, 
in the proposed RL algorithm,  the agent observes the amount of data most recently downloaded from the SBSs, as well as the amount of average cache space occupied by each file. Upon observation of a file request, the agent takes an action, i.e., makes a decision on the caching policy \eqref{eq:policy} (using \eqref{eq:synchpol}). Note that the action results in an amount of data downloaded from the SBSs and an average cache usage at the time of the next request for that particular file. This information is observed by the agent already at the next state transition, together with the file requested next in the sequence of requests. Furthermore, the action results in an amount of data sent to update all caches. As the aim is to determine caching policies that maximize \eqref{eq:maximization}, we set the reward that the agent receives based on the amount of data downloaded from the caches and a penalty for the amount of data sent to update the caches as well as for deviating from the cache size constraint. We note that considering a long-term average cache size constraint as in \cite{Goseling2019,Pedersen2020} would make the training of the agent a painfully slow process. To circumvent this shortcoming, we consider the average occupied cache space due to the most recent action and inter-request time instead. The goal is for the agent to learn to consistently take optimal actions to maximize \eqref{eq:maximization} while adhering to the cache size constraint.

\subsection{State}
\label{sec:state}
We model the (real) request times $t_0, t_1, \ldots$ as the discrete time-steps of the MDP. To be consistent with the definition of MDPs in Section~\ref{sec:prel}, we refer to the request times by their subindex, i.e., we consider the request \emph{time-steps}
\begin{equation}
	t = 0, 1, 2, \ldots.
\end{equation}
Unless stated otherwise, we will refer to $t$ as a time-step, $t+1$ as the next time-step, etc., for the remainder of the paper. Fig.~\ref{fig:map} shows an illustration of requests in real time and modeled as discrete MDP time-steps. 

Each state $S_t$ contains information about the file requested at time-step $t$, denoted by $f^{(t)}$. Furthermore, the state also contains information about the amount of each file cached at the SBSs, denoted by $\mu_f^{(t)}$, $f = 1, 2, \ldots, F$, and the average cache usage of each file, denoted by $\mubar_f^{(t)}$, $f = 1, 2, \ldots, F$. We defer a further explanation of $\mu_f^{(t)}$ and $\mubar_f^{(t)}$ to Section~\ref{sec:nextstate}. The state is hence defined as
\begin{equation}
\label{eq:state}
	S_t \triangleq \left[ f^{(t)}, \boldmu^{(t)}, \boldmubar^{(t)} \right],
\end{equation}
with
\begin{equation}
	\boldmu^{(t)} = \left[ \mu_1^{(t)}, \mu_2^{(t)}, \ldots, \mu_F^{(t)} \right]
\end{equation}
and
\begin{equation}
	\boldmubar^{(t)} = \left[ \mubar_1^{(t)}, \mubar_2^{(t)}, \ldots, \mubar_F^{(t)} \right].
\end{equation}

\subsection{Action}
We define the action $A_t$ to be the caching policies \eqref{eq:policy} for the requested file at time-step $t$, $f^{(t)}$, under synchronously updated caches, i.e., using \eqref{eq:synchpol},
\begin{equation}
\label{eq:action}
	A_t = \left[ x_{f^{(t)}}^{(0)}, x_{f^{(t)}}^{(1)}, \ldots, x_{f^{(t)}}^{(K)} \right].
\end{equation}

\subsection{Next State}
\label{sec:nextstate}
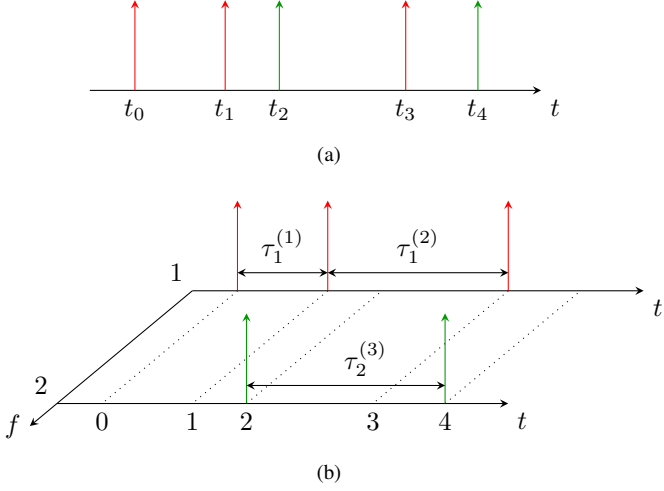
\begin{figure}[t!]
	\begin{subfigure}[t]{\columnwidth}
		\centering
		\begin{tikzpicture}[>=stealth, scale=1.2]
			\draw[->] (0, 0) to (5, 0) node[below right]{$t$};
			
			\draw[red, ->] (0.5, 0) node[black, below]{$t_0$} to +(0, 1);
			\draw[red, ->] (1.5, 0) node(r2)[black, below]{$t_1$} to +(0, 1);
			\draw[green, ->] (2.1, 0) node[black, below]{$t_2$} to +(0, 1);
			\draw[red, ->] (3.5, 0) node(r4)[black, below]{$t_3$} to +(0, 1);
			\draw[green, ->] (4.3, 0) node[black, below]{$t_4$} to +(0, 1);
			
		\end{tikzpicture}
		\caption{}
	\end{subfigure}
	\par\bigskip
	\begin{subfigure}[t]{\columnwidth}
		\centering
		\begin{tikzpicture}[>=stealth, scale=1.2]
			\draw[->] (1.5, 1.25) node[above left] {$1$} to +(5, 0) node[below right]{$t$};
			\draw[->] (1.5, 1.25) to +(-1.8, -1.5) node[left] {$f$};
			\draw[red, ->] (2, 1.25) node(r1){} to +(0, 1);
			\draw[red, ->] (3, 1.25) node(r2){} to +(0, 1);
			\draw[red, ->] (5, 1.25) node(r4){} to +(0, 1);
			
			\draw[dotted] (r1.base) to +(-1.5, -1.25) node[below]{$0$};
			\draw[dotted] (r2.base) to +(-1.5, -1.25) node[below]{$1$};
			\draw[dotted] (r4.base) to +(-1.5, -1.25) node[below]{$3$};
			
			\draw[->] (0, 0) node[above left] {$2$} to +(5, 0) node[below right]{$t$};
			\draw[green, ->] (2.1, 0) node(r3){} to +(0, 1);
			\draw[green, ->] (4.3, 0) node(r5){} to +(0, 1);
			
			\draw[dotted] (r3.base) node[below]{$2$} to +(1.5, 1.25);
			\draw[dotted] (r5.base) node[below]{$4$} to +(1.5, 1.25);
			
			\draw[<->] ($(r1.base) + (0, 0.2)$) -- ($(r2.base) + (0, 0.2)$) node[midway, above]{$\tau_1^{(1)}$};
			\draw[<->] ($(r2.base) + (0, 0.2)$) -- ($(r4.base) + (0, 0.2)$) node[midway, above]{$\tau_1^{(2)}$};
			\draw[<->] ($(r3.base) + (0, 0.2)$) -- ($(r5.base) + (0, 0.2)$) node[pos=0.6, above]{$\tau_2^{(3)}$};
		\end{tikzpicture}
		\caption{}
	\end{subfigure}
	\caption{An example of the mapping of file requests from real time (a) to discrete time-steps (b). Red and green arrows correspond to requests for file $1$ and $2$, respectively, e.g., $f^{(0)} = 1$ and $f^{(4)}= 2$, etc..}
	\label{fig:map}
	\vspace{-2ex}
\end{figure}

At time-step $t$, file $f^{(t)}$ is requested. Assume that the same file is requested again at time-step $t + n$, for some $n = 1, 2, \ldots$. The (real) time between the two consecutive requests for file $f^{(t)}$ is the inter-request time defined in Section~\ref{sec:policy}. Even though this time is not observed until time-step $t + n$, we make it observable to the agent non-causally at time-step $t+1$ and denote it by $\tau_{f^{(t)}}^{(t+1)}$. The reason for making the inter-request time observable at time-step $t+1$ is that it will be used in the computation of the reward $R_{t+1}$ (explained in Section~\ref{sec:reward}), which is necessarily a function of $S_t$, $A_t$, and $S_{t+1}$ (defined in \eqref{eq:nextstate}) \cite[Section~3.1]{Sutton2018}. Furthermore, making the inter-request time $\tau_{f^{(t)}}^{(t+1)}$ observable at time-step $t+1$ has no implications as the agent took already action $A_t$ at time-step $t$. The inter-request times and the time-steps when they are observed are highlighted in Fig.~\ref{fig:map}(b).

The inter-request time $\tau_{f^{(t)}}^{(t+1)}$ corresponds to a request for file $f^{(t)}$ appearing in a time-slot $j = 0, 1, \ldots, K$ of the previously selected caching policy (see Section~\ref{sec:policy}), i.e., action $A_t$. Specifically, the inter-request time equates to a request in time-slot
\begin{equation}
\label{eq:ell}
	\ell = \min \left\{ \floor{ \frac{\tau_{f^{(t)}}^{(t+1)}}{T}}, K \right\},
\end{equation}
where we omit the dependence of $\ell$ on $t$ for ease of exposition. Furthermore, a request in time-slot $\ell$ results in a fraction $x_{f^{(t)}}^{(\ell)}$ being cached at all SBSs, as per the caching policy. The elements of the vector $\boldmu^{(t+1)}$ are updated as follows,
\begin{equation}
\label{eq:munext}
\mu_f^{(t+1)} = \begin{cases}
	x_{f^{(t)}}^{(\ell)}, & \text{if}~f = f^{(t)},\\
	\mu_f^{(t)}, & \text{otherwise},
\end{cases}
\end{equation}
i.e., the amount of file $f^{(t)}$ available to download due to the inter-request time $\tau_{f^{(t)}}^{(t+1)}$ and action $A_t$ is updated, and the information regarding the cached amount of other files is kept the same. Using \eqref{eq:policy}, the average memory used to cache file $f^{(t)}$ due to action $A_t$ and inter-request time $\tau_{f^{(t)}}^{(t+1)}$ is
\begin{align}
	\xbar_{f^{(t)}} & = \frac{1}{\tau_{f^{(t)}}^{(t+1)}} \int_0^{\tau_{f^{(t)}}^{(t+1)}} x_{f^{(t)}}(s)~\d s\\
		& = \frac{1}{\tau_{f^{(t)}}^{(t+1)}} \left( T \sum_{j=0}^{\ell - 1} x_{f^{(t)}}^{(j)} + \left( \tau_{f^{(t)}}^{(t+1)} - \ell T \right) x_{f^{(t)}}^{(\ell)} \right), \label{eq:xbar}
\end{align}
where we utilize the empty sum convention, i.e.,
\begin{equation}
	\sum_{j=0}^{\ell - 1} x_{f^{(t)}}^{(j)} = 0,
\end{equation}
if $\ell = 0$. Similar to the update of the amount cached, the vector $\boldmubar^{(t+1)}$ is updated with the latest estimate of the long-term average cache occupancy of file $f^{(t)}$, i.e., 
\begin{equation}
\label{eq:mubarnext}
\mubar_f^{(t+1)} = \begin{cases}
	\xbar_{f^{(t)}}, & \text{if}~f = f^{(t)},\\
	\mubar_f^{(t)}, & \text{otherwise}.
\end{cases}
\end{equation}
The next state $S_{t+1}$ includes the file requested at time-step $t+1$, $f^{(t+1)}$, as well as the cached amounts and estimated cache occupancy of each file, i.e.,
\begin{equation}
\label{eq:nextstate}
	S_{t+1} = \left[ f^{(t+1)}, \boldmu^{(t+1)}, \boldmubar^{(t+1)} \right].
\end{equation}
The following example illustrates the states and actions of the MDP. Specifically, we demonstrate how to obtain $x_{f^{(t)}}^{(\ell)}$ and $\xbar_{f^{(t)}}$.

\begin{figure}[t!]
	\centering
	\begin{tikzpicture}[>=stealth, scale=1.2]
		\draw[->] (1.5, 1.25) to +(5, 0) node[below right]{$t$};
		\draw[->] (1.5, 1.25) node[above left] {$1$} to +(-1.8, -1.5) node[left]{$f$};
		\draw[red, ->] (2, 1.25) node(r1){} to +(0, 1);
		\draw[red, ->] (3, 1.25) node(r2){} to +(0, 1);
		\draw[red, ->] (5, 1.25) node(r4){} to +(0, 1) node[black, above]{$\B^{(2)}$};
		
		\draw[dotted] (r1.base) to +(-1.5, -1.25) node[below]{$0$};
		\draw[dotted] (r2.base) to +(-1.5, -1.25) node[below]{$1$};
		\draw[dotted] (r4.base) to +(-1.5, -1.25) node[below]{$3$};
		
		\draw[->] (0, 0) node[above left] {$2$} to +(5, 0) node[below right]{$t$};
		\draw[green, ->] (2.1, 0) node(r3){} to +(0, 1);
		
		\draw[dotted] (r3.base)  node[below]{$2$} to +(1.5, 1.25);
		
		\draw ($(r2.base) + (0, 0.8)$) node[left]{$x_{1}^{(0)}$} -| ++(1.2, -0.4) node[left]{$x_{1}^{(1)}$} -- ++(1.2, 0) -- ++(0, -0.4) node[below]{$x_{1}^{(2)}$};
		\draw[<->] ($(r2.base) + (0, 0.1)$) -- ($(r4.base) + (0, 0.1)$) node[midway, below = 2pt]{$\tau_1^{(2)}$};
		
		\node (mu) at (5, 1.65) {};
		\node (mulabel) at ($(mu) + (1, 0.2)$) {$\mu_1^{(2)}$};
		\draw (mulabel) -- (mu.base);
		
		\draw[dashed] ($(r2.base) + (0, 0.65)$) -- ($(r4.base) + (0, 0.65)$) node[pos=0.9](mubar){};
		\node (mubarlabel) at ($(mubar) + (1, 0.5)$) {$\mubar_1^{(2)}$};
		\draw[bend left] (mubarlabel) -- (mubar.base);
	\end{tikzpicture}
	\caption{The states and actions of the MDP used in Example~\ref{ex:stateaction}.}
	\label{fig:states}
	\vspace{-2ex}
\end{figure}
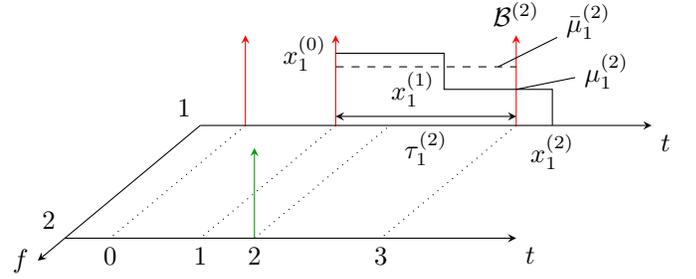

\begin{example}
\label{ex:stateaction}
	Consider a library of $F = 2$ files, update period $T = 1$, and $K = 2$ cache updates for which the current state and action at time-step $t = 1$, as well as the next state at time-step $t+1 = 2$, is illustrated in Fig.~\ref{fig:states}. For the file $f^{(1)} = 1$ requested at time-step $t = 1$, we assume that the agent takes an action, i.e., makes a decision on the caching policy (see \eqref{eq:action}),
	\begin{equation}
		A_1 = \left[ x_1^{(0)}, x_1^{(1)}, x_1^{(2)} \right] = [1, 0.5, 0].
	\end{equation}
	In the example, the next request for file $f^{(1)} = 1$ appears at time-step $t + 2 = 3$. However, the inter-request time, assumed to be $\tau_1^{(2)} = 1.7$, is observable at time-step $t + 1 = 2$. Using \eqref{eq:ell}, we see that the file is requested in time-slot
	\begin{equation}
		\ell = \min \left\{ \floor{ \frac{\tau_1^{(2)}}{T}}, K \right\} = \min \left\{ \floor{ \frac{1.7}{1}}, 2 \right\} = 1
	\end{equation}
	of the selected caching policy. Hence,
	\begin{equation}
		x_1^{(\ell)} = x_1^{(1)} = 0.5
	\end{equation}
	is the element $\mu_1^{(2)}$ of $\boldmu^{(2)}$. Furthermore, using \eqref{eq:xbar},
	\begin{align}
		\xbar_1 & = \frac{1}{\tau_1^{(2)}} \left( T \sum_{j=0}^{\ell - 1} x_1^{(j)} + \left( \tau_1^{(2)} - \ell T \right) x_1^{(\ell)} \right)\\
			& = \frac{1}{1.7} \left( x_1^{(0)} + 0.7 x_1^{(1)} \right) \approx 0.79,
	\end{align}
	which is hence the element $\mubar_1^{(2)}$ of $\boldmubar^{(2)}$. The file requested at the next time-step (i.e., $t+1 = 2$) is $f^{(2)} = 2$. Using \eqref{eq:nextstate}, the next state is hence
	\begin{equation}
		S_{2} = \left[ 2, \boldmu^{(2)}, \boldmubar^{(2)} \right].
	\end{equation}
\end{example}

Similar to how the inter-request times are made available to the agent non-causally at the next time-step $t+1$, we use the same convention for the set of SBSs within communication range of the user placing the next request for file $f^{(t)}$. If there is a subsequent request for file $f^{(t)}$ at time-step $t + n$, for some $n = 0, 1, \ldots$, we make the set of SBSs within communication range of the user placing the request at time-step $t + n$ observable to the agent non-causally at the next time-step $t +1$ and denote the set by $\B^{(t+1)}$ (see Fig.~\ref{fig:states} for an illustration assuming $t+1 = 2$). However, note that we do not include $\B^{(t+1)}$ as state information in $S_{t+1}$, as that not doing so we have observed improves the performance during training. Hence, the MDP is strictly a partially observable MDP \cite[Chapter~17]{Sutton2018}.

\subsection{Reward}
\label{sec:reward}
At the next request for file $f^{(t)}$, $|\B^{(t+1)}|$ SBSs are within communication range of the user placing that particular request. Using \eqref{eq:sbsdata} and \eqref{eq:synch}, the amount of file $f^{(t)}$ downloaded from SBSs at the next request is hence
\begin{equation}
\label{eq:Rdata}
	\Rsbs = \min \left\{ \left| \B^{(t+1)} \right| \mu_{f^{(t)}}^{(t+1)}, 1 \right\}.
\end{equation}
The amount of data sent to update all caches, as a consequence of action $A_t$ and a request in time-slot $\ell$ (see \eqref{eq:ell}), is
\begin{align}
	\Rupd & = B \bigg( \max \left\{ x_{f^{(t)}}^{(0)} - \mu_{f^{(t)}}^{(t)}, 0 \right\} \nonumber \\
		& \hspace{3ex} + \sum_{j = 1}^\ell \max \left\{ x_{f^{(t)}}^{(j)} - x_{f^{(t)}}^{(j-1)}, 0 \right\} \bigg), \label{eq:Rupd}
\end{align}
where \eqref{eq:Rupd} accounts for any increase in the amount cached over time-slots. As explained in Section~\ref{sec:state}, each request corresponds to a time-step and thereby a (real) time. Therefore, the unit of \eqref{eq:Rdata} and \eqref{eq:Rupd} is an amount of data per time, i.e., data rate, by definition. Hence, maximizing $\Rsbs - \updcost \Rupd$ corresponds to minimizing the network load \eqref{eq:netwload}, i.e., maximizing \eqref{eq:maximization}.

A long-term cache memory constraint is considered in \cite{Pedersen2020}, which is not possible for our RL approach. To see why, consider, e.g., a drastic overuse of the available cache space at some point during the training. The agent would consequently have to under-utilize the available cache space for numerous training episodes to compensate for the previous overuse without learning the optimal caching policy, which would drastically slow down training. Instead, we penalize the deviation in memory usage from the cache storage capacity using the last available estimates of cache memory usage, i.e.,
\begin{equation}
\label{eq:Rmem}
	\Rmem = \left| \sum_{f=1}^F \mubar_f^{(t+1)} - C \right|.
\end{equation}
Consequently, we set our reward to
\begin{equation}
\label{eq:reward}
	R_{t+1} = \Rsbs - \updcost \Rupd - \Rmem.
\end{equation}

\subsection{Termination}
\label{sec:term}
We assume that the episode, defined as a number of requests, terminates at some point. Specifically, we assume that the episode terminates when the request at time-step $t+1$ is the last request for file $f^{(t+1)}$ since no inter-request time exists beyond this request for the particular file. Finally, to distinguish the RL approach in this section to the MARL framework presented in the following section, we refer to the former as \emph{single-agent} RL (SARL).

\section{MARL-Based Coded Caching for\\Asynchronous Cache Updates}
\label{sec:asynch}
For a spatially heterogenous request process, synchronously updated caches may not be optimal. Hence, in this section, we lift this assumption, i.e., the constraint in \eqref{eq:synchpol}. To address asynchronous updates, we propose a decentralized approach where each SBS is an independent agent, taking independent decisions on the caching policies \eqref{eq:policy}. We formulate this scenario as an MARL problem, where the SBSs are independent agents, able to autonomously adapt to a spatially heterogenous request process, something which is impossible under the centralized approach with synchronously updated caches in Section~\ref{sec:synch}. In the following, we define the state, action, and reward as observed by one agent $\refagent \in \{1, 2, \ldots, B\}$, henceforth referred to as the reference agent. We omit labeling the state, action, and reward with the reference agent index for readability.

\subsection{State}
Let the request times represent MDP time-steps, analogously to the single-agent case (see Section~\ref{sec:state}), with one key exception: only the requests within communication range of the reference SBS (agent) are considered. In addition to the state information provided to the SARL algorithm, shown in \eqref{eq:state}, the reference agent also observes the estimated amount of a file downloaded from other SBSs, denoted by $\tilde\boldmu^{(t)}$. Specifically,
\begin{equation}
	S_t = \left[ f^{(t)}, \boldmu^{(t)}, \boldmubar^{(t)}, \tilde\boldmu^{(t)} \right],
\end{equation}
where
\begin{equation}
	\tilde\boldmu^{(t)} = \left[ \tilde\mu_{1}^{(t)}, \ldots, \tilde\mu_{\refagent-1}^{(t)}, \tilde\mu_{\refagent+1}^{(t)}, \ldots, \tilde\mu_{B}^{(t)}\right].
\end{equation}
We defer a further explanation of the state vector, in particular the vector $\tilde\boldmu^{(t)}$, until Section~\ref{sec:marlnextstate}.

\subsection{Action}
Using the MARL scheme, the reference agent chooses a caching policy in a similar fashion to the single-agent case, i.e., the action is given by \eqref{eq:action}.

\subsection{Next State}
\label{sec:marlnextstate}
The information regarding the amount of data cached locally as well as the average cache occupancy is computed according to \eqref{eq:munext} and \eqref{eq:mubarnext}, using \eqref{eq:ell} and \eqref{eq:xbar}, respectively. From the perspective of the reference agent, $\B^{(t+1)}$ is the set of SBSs within communication range of a user placing the next request for file $f^{(t)}$. Note that the definition of the set $\B^{(t+1)}$ is the multi-agent analog, i.e., observations are independent to each agent, to the single-agent case in Section~\ref{sec:nextstate}. The reference agent receives information about the amount of file $f^{(t)}$ cached by SBSs $b \in \B^{(t+1)} \backslash \refagent$ at the time of the next request, denoted by $\tilde\mu_b^{(t+1)},~b \in \B^{(t+1)} \backslash \refagent$. The cached amount is a currently available estimate since agents $\B^{(t+1)} \backslash \refagent$ are potentially involved in state transitions not involving $\refagent$ until the next request for $f^{(t)}$. Hence, we highlight this fact with tilde. Furthermore, transmitting estimates is reasonable if inter-request times $\tau_{f^{(t)}}^{(t+1)}$ are approximately independent and identically distributed (i.i.d.). The next state is
\begin{equation}
	S_{t+1} = \left[ f^{(t+1)}, \boldmu^{(t+1)}, \boldmubar^{(t+1)}, \tilde\boldmu^{(t+1)} \right].
\end{equation}

\subsection{Reward}
At the time of the next request for file $f^{(t)}$, an estimated amount
\begin{equation}
\label{eq:Rdatamarl}
	\Rsbs = \min \left\{  \mu_{f^{(t)}}^{(t+1)} + \sum_{b \in \B^{(t+1)} \backslash \refagent} \tilde{\mu}_b^{(t+1)} , 1\right\}
\end{equation}
is downloaded from SBSs. The amount of data transmitted to update the cache of the reference agent is
\begin{align}
	\Rupd & = \bigg( \max \left\{ x_{f^{(t)}}^{(0)} - \mu_{f^{(t)}}^{(t)}, 0 \right\} \nonumber \\
		& \hspace{3ex} + \sum_{j = 1}^\ell \max \left\{ x_{f^{(t)}}^{(j)} - x_{f^{(t)}}^{(j-1)}, 0 \right\} \bigg),
\end{align}
only distinguished from \eqref{eq:Rupd} by the scaling of the number of SBSs $B$. For the MARL scheme, the penalty for deviations from the cache size constraint is equivalent to \eqref{eq:Rmem} and the reward $R_{t+1}$ is given by \eqref{eq:reward}.

\subsection{Termination}
Similar to the single-agent case explained in the previous section, an episode terminates when the request at time-step $t+1$ is the last request for file $f^{(t+1)}$. However, we allow terminated agents to keep passing estimates of the cached amount to other non-terminated agents as these are needed to compute \eqref{eq:Rdatamarl}.

\section{Numerical Results}
\label{sec:results}
For the numerical results, we consider the scenario where SBSs are placed on a grid, which is the same model used in \cite{Bioglio2015}, with distance between the SBSs equal to $1$. We assume a communication range $1/\sqrt{2} \le r \le 1$ and focus on a single square in the grid and refer to it as the \emph{area}. We assume that users in the area request files with times between requests i.i.d. according to a Weibull distribution, which has been shown to accurately capture the inter-request times of video content \cite{Costa2004}, i.e., $\tau_{f^{(t)}}^{(t+1)} = \tau_f \sim \text{Weibull}(k_f, \lambda_f)$, where $k_f$, $0 < k_f \le 1$, and $\lambda_f$ are the shape and scale parameters of the distribution, respectively. Note that the dependence of $\tau_f$ on $t$ is removed due to the i.i.d. assumption. The expected inter-request time is
\begin{equation}
	\omega_f^{-1} = \E[\tau_f] = \lambda_f \Gamma(1+k_f^{-1}),
\end{equation}
where $\Gamma(\cdot)$ is the Gamma function. We furthermore assume that $p_f$ is the Zipf probability mass function with parameter $\alpha$, i.e.,
\begin{equation}
	p_f = \frac{1/f^\alpha}{\sum_{j=1}^F 1/j^\alpha},~\alpha \ge 0.
\end{equation}

Unless stated otherwise, we will assume the following setup for the remainder of this section. The library holds $F = 20$ files, each of normalized size $1$. We set the Weibull shape $k_f = k = 0.6$, which describes a quite bursty request process and is within the range specified in \cite{Costa2004}. Also, we set $\alpha = 0.7$, which has been shown to accurately capture the popularity of YouTube videos \cite{Cheng2008}. We furthermore assume an aggregate request rate $\omega = 100$. Finally, we consider $K = 2$ cache updates and an update period $T = 0.5$. The details regarding the DDPG algorithm configuration are provided in the appendix.

\begin{figure}[t!]
	\centering
	\includegraphics[width=\columnwidth]{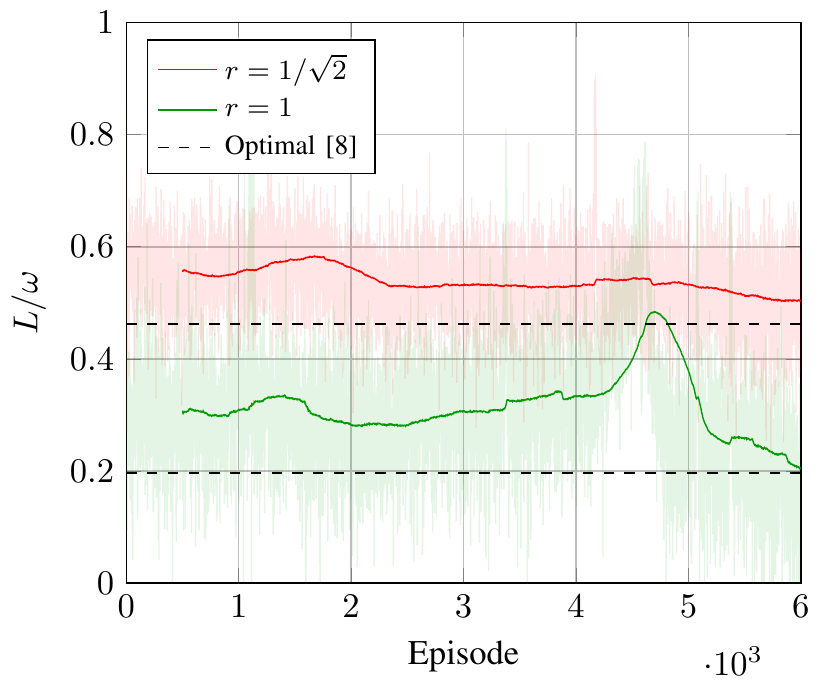}
	\caption{The normalized network load versus the training episode for $\updcost = 0.05$. The thick curves correspond to $500$ episode moving averages.}
	\label{fig:train}
	\vspace{-2ex}
\end{figure}

\subsection{Homogenous Request Process}
We first assume that the request process is spatially homogeneous, i.e., the users requesting files are assumed to be uniformly distributed in the area. Hence, we consider the case where all caches are synchronously updated, i.e., all SBSs cache always the same amount of each file (although corresponding to unique coded packets). We use the DDPG algorithm to solve the MDP proposed in Section~\ref{sec:synch}, i.e., to obtain caching policies \eqref{eq:policy} minimizing the network load \eqref{eq:netwload}, and compare these to the optimal caching policies in \cite{Pedersen2020}, which assume perfect knowledge of the request statistics and SBS distribution. We consider a communication range $r = 1/\sqrt{2}$ and $r = 1$, corresponding to an expected number of SBSs within communication range of a user equal to approximately $1.57$ and $3.14$, respectively. We also assume an average cache size constraint $C = 4$ and a cache update cost $\updcost = 0.05$, i.e., the update cost is $5\%$ of the  cost $1-\sbscost = 1$ (see \eqref{eq:relcost}). For this scenario, Fig.~\ref{fig:train} shows the network load \eqref{eq:netwload} achieved by the RL approach prosed in Section~\ref{sec:synch}, normalized by the aggregate request rate $\omega$, when training the DDPG algorithm, where the thick curves are $500$ episode moving averages. For the example shown in the figure, we train the algorithm for $5 \cdot 10^3$ episodes with fixed exploration noise variance $\sigma^2$. The variance is decreased linearly to zero for the last $10^3$ episodes. Simulating the caching policies without exploration noise and NN updates (not shown in the figure), we observe the network load $L/\omega = 0.511$ and $L/\omega = 0.203$ for $r = 1/\sqrt{2}$ and $r = 1$, respectively. The corresponding network load achieved by optimized caching policies found using \cite{Pedersen2020} is $L/\omega = 0.462$ and $L/\omega = 0.197$. Hence, the caching policies obtained through RL attain a network load within $10.6\%$ and $3.0\%$ from the optimal network load, respectively.

\begin{figure}[t!]
	\centering
	\includegraphics[width=\columnwidth]{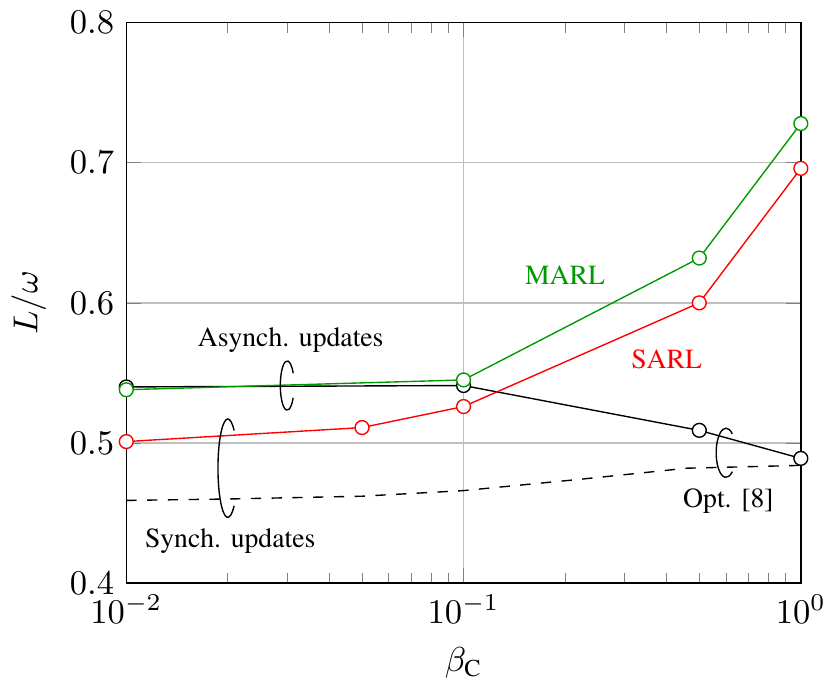}
	\caption{The normalized network load versus the cache update cost using caching policies obtained through the RL frameworks and through optimization under a homogenous (in space) request process.}
	\label{fig:updcost}
	\vspace{-2ex}
\end{figure}

Under synchronous cache updates, all caches are updated regardless of which SBSs are within communication range of the user placing the request. Depending on the update cost, updating all SBSs at each update could have a big impact on the network load. Hence, an asynchronous cache update policy might potentially achieve lower network load, as in this case the SBSs are instead updated individually, which reduces the update cost. To gain some insight, in Fig.~\ref{fig:updcost} we plot the normalized network load $L/\omega$ achieved by the SARL framework proposed in Section~\ref{sec:synch}\textemdash which assumes synchronous updates\textemdash and the MARL approach proposed in Section~\ref{sec:asynch}\textemdash which by nature captures asynchronous updates\textemdash versus the cache update cost $\updcost$. We assume that $r = 1/\sqrt{2}$ and $C = 4$. Note that the training curve for SARL is provided in Fig.~\ref{fig:train} for $\updcost = 0.05$. As a reference, we also plot the minimum normalized network load obtained through the optimization of caching policies in \cite{Pedersen2020}, which assumes perfect knowledge of request statistics and SBS distribution, and synchronous cache updates (dashed curve). We remark that the optimization framework \cite{Pedersen2020} cannot be extended to asynchronous updates, as the optimization problem becomes intractable. We furthermore benchmark the $L/\omega$ achieved by our proposed RL algorithms to the optimized caching policies using the framework in \cite{Pedersen2020}, i.e., for synchronous updates, simulated under an asynchronous cache update scenario (solid black curve). We observe that the SARL algorithm outperforms the MARL algorithm and that the caching policies optimized using \cite{Pedersen2020} yield the lowest network load for all values of $\updcost$. Caching policies obtained through MARL are on par with caching policies optimized using \cite{Pedersen2020} and simulated under asynchronous updates in terms of network load for $\updcost \le 10^{-1}$. As $\updcost \to 1$, the optimal caching policy tends to static caching \cite{Pedersen2020}, i.e.,
\begin{equation}
	x_f^{(0)} = x_f^{(1)} = \ldots = x_f^{(K)},~\text{for all}~f.
\end{equation}
The RL approaches proposed in this paper fail to find static caching, which is why the performance suffers for $\updcost > 10^{-1}$. While we cannot prove that synchronous updates minimize the network load under a spatially homogenous request process and, in fact, this research problem remains open, the results in Fig.~\ref{fig:updcost} suggest that synchronized cache updates is optimal, despite the update cost. This is because the update cost has a smaller impact on the network load than the fact that the spatially homogeneous request process has memory. The probability of requests in short succession is large, but subsequent requests are equally likely to come from users within communication range of other SBSs, hence it is beneficial to push updates to all the caches.

\subsection{Heterogenous Request Process}
To simulate a request process that is heterogenous in space, we devise the following setup. We divide the file library in $B = 4$ classes by the operation $f \bmod B$. The probability that the user requesting file $f$ at (real) time $t$ is within communication range of a particular SBS, i.e., $b \in \B^{(t)}$, given that a particular file is requested, is
\begin{equation}
	\zeta \triangleq \Pr \left\{ b \in \B^{(t)}~|~f \bmod B = (b-1) \right\}.
\end{equation}
Apart from the conditioning that a user requesting a particular file is within communication range of a particular SBS, the users are assumed to be uniformly distributed. Note that, for $r = 1/\sqrt{2}$, $\zeta = \pi/8 \approx 0.39$ corresponds to users being uniformly distributed in the whole area. The request process is less homogenous the more $\zeta$ deviates from this value. 

\begin{figure}[t!]
\centering
	\includegraphics[width=\columnwidth]{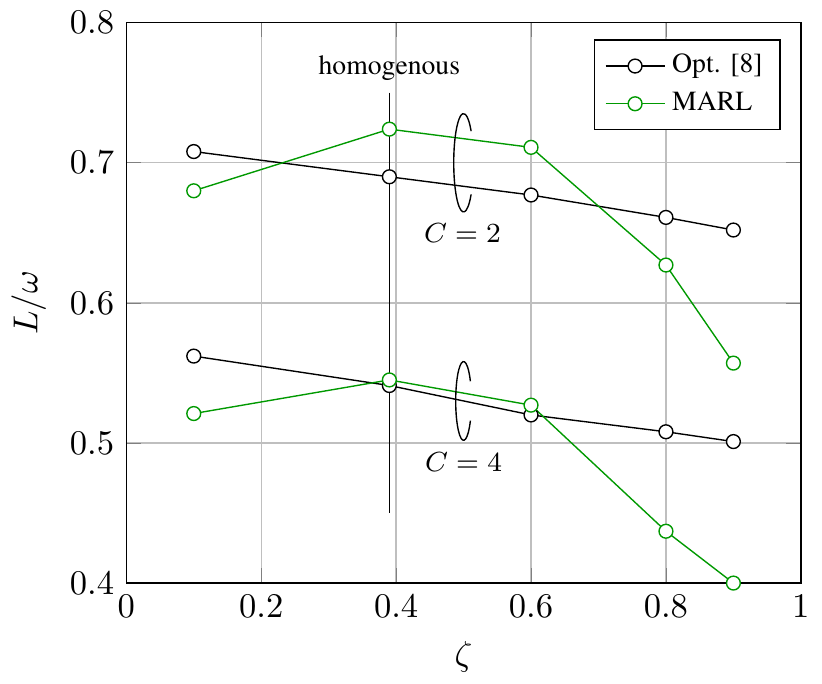}
	\caption{The normalized network load versus the probability that a user requesting a file is within communication range of one of the SBSs, i.e., for the scenario of a spatially heterogenous request process, assuming $\updcost = 0.1$. $\zeta = \pi/8$, corresponding to a homogenous request process, is indicated by the vertical line.}
	\label{fig:nonuniform}
	\vspace{-2ex}
\end{figure}

Fig.~\ref{fig:nonuniform} shows the normalized network load when using caching policies obtained using the MARL framework in Section~\ref{sec:asynch}, assuming an update cost $\updcost = 0.1$ as well as a cache size constraint $C = 2$ and $C = 4$. We include the $L/\omega$ achieved by caching policies optimized using \cite{Pedersen2020}, which necessarily assume a homogenous request process as well as perfect knowledge of request statistics and SBS distribution, simulated for the scenario of the heterogenous request process.

The MARL caching policies reduce the network load compared to when using the optimized caching policies the more non-uniformly distributed the users requesting particular files become. The MARL caching policy decreases the network load compared to optimized policies for $\zeta < \pi/8$, i.e., for a smaller probability of a user requesting a file belonging to a class associated with a particular SBS being within communication range of the particular SBS. However, the gains in terms of network load are even greater for an increase in the probability, i.e., when $\zeta > \pi/8$. For example, the MARL caching policy reduces the network load by $20\%$ for $\zeta = 0.9$ and $C = 4$.

\section{Conclusion}
We proposed a reinforcement learning (RL) approach to learn soft time-to-live caching policies to minimize the network load for a scenario where content is encoded using maximum distance separable codes and cached in a distributed fashion across several small base stations (SBSs). Specifically, we proposed a single-agent RL algorithm which learns caching policies under synchronous cache updates, suitable to spatially homogenous request processes. The algorithm learns caching policies achieving almost the same network load as optimized caching policies without requiring any prior knowledge of request statistics or SBS distribution, which is necessary in order to find the optimized policies. For homogenous (in space) request processes, our results suggest that synchronizing the cache updates reduces the network load, although the problem remains open. We furthermore proposed a multi-agent RL (MARL) approach, where the SBSs are modeled as independent agents, which can autonomously learn caching policies, i.e., the proposed MARL approach captures asynchronous cache updates.  For  spatially heterogenous request processes, we demonstrated that the caching policies learned by our MARL algorithm reduce the network load significantly as compared to optimized caching policies simulated under asynchronous cache updates, where the optimization assumes both synchronously updated caches and perfect knowledge of request statistics and SBS distribution.

\appendix
We consider the following configuration for the DDPG algorithm \cite{Lillicrap2019}. We encode the requested file $f^{(t)}$ as a one-hot vector, to make it easier for the NNs to distinguish the requested file. The actor and critic networks comprise two hidden layers with $64$ neurons in each layer. We use batch norm and rectified linear unit activation functions after each layer. The outputs of the actor networks, i.e., the actions, are sigmoid activation functions, to which we add zero-mean Gaussian noise with variance $\sigma^2$. We use the Adam optimizer \cite{Kingma2017} for stochastic gradient descent. All parameter values are provided in Table~\ref{tab:params}. For the SARL case, we decrease the exploration noise, i.e., $\sigma^2$, linearly, akin to simulated annealing. For the MARL case, we maintain exploration throughout training. During testing, both exploration and NN weight updating is turned off.

\ctable[
	caption = {The parameter values used for the DDPG algorithm.},
	label={tab:params},
	width=\columnwidth
]{Xl}{}{
	\FL 
		Parameter & Value
	\ML 
		Actor network learning rate & $10^{-4}$\\
		Critic network learning rate & $10^{-3}$\\
		Target network update $\rho$ & $0.999$\\
		Memory buffer size $|\buffer|$ & $10^6$\\
		Batch size $|\replay|$ & $64$\\
		Discount rate $\gamma$ & $0.99$\\
		Exploration noise variance $\sigma^2$ & 0.01
	\LL 
}

\bibliographystyle{IEEEtran}
\bibliography{confs-jrnls,IEEEabrv,library}

\end{document}